\documentclass{ifacconf}

\usepackage{}        
\usepackage{natbib,amsmath,amssymb,graphicx,subcaption,tikz-cd,algorithm}
\usepackage[italicComments=true,noEnd=false,indLines=true]{algpseudocodex}
\newcommand{\RR}{\mathbb{R}}
\newcommand{\XX}{\mathbb{X}}
\newcommand{\D}{\mathcal{D}}
\renewcommand{\H}{\mathcal{H}}
\newcommand{\K}{\mathcal{K}}

\newcommand{\M}{\mathcal{M}}
\newcommand{\G}{\mathcal{G}}
\newcommand{\U}{\mathcal{U}}
\begin{document}
\begin{frontmatter}

\title{Online learning of Koopman operator using streaming data from different dynamical regimes \thanksref{footnoteinfo}} 

\thanks[footnoteinfo]{Sponsor and financial support acknowledgment
goes here. Paper titles should be written in uppercase and lowercase
letters, not all uppercase.}

\author[First]{Kartik Loya} 
\author[First]{Phanindra Tallapragada} 

\address[First]{Department of Mechanical Engineering, Clemson University, Clemson, SC.(e-mail: kloya@clemson.edu, ptallap@clemson.edu)}

\begin{abstract}
The paper presents a framework for online learning of the Koopman operator using streaming data. Many complex systems for which data-driven modeling and control are sought provide streaming sensor data, the abundance of which can present computational challenges but cannot be ignored. Streaming data can intermittently sample dynamically different regimes or rare events which could be critical to model and control. Using ideas from subspace identification, we present a method where the Grassmannian distance between the subspace of an extended observability matrix and the streaming segment of data is used to assess the `novelty' of the data. If this distance is above a threshold, it is added to an archive and the Koopman operator is updated if not it is discarded. Therefore, our method identifies data from segments of trajectories of a dynamical system that are from different dynamical regimes,  prioritizes minimizing the amount of data needed in updating the Koopman model and furthermore reduces the number of basis functions by learning them adaptively. Therefore, by dynamically adjusting the amount of data used and learning basis functions, our method optimizes the model's accuracy and the system order.

\end{abstract}

\begin{keyword}
Koopman operator, online learning, recursive subspace identification, Grassmannian distance
\end{keyword}

\end{frontmatter}

\section{INTRODUCTION}
The increasing ubiquity of sensors has led to an abundance of data of measurements of complex systems, which, combined with improved computational resources, has led to an interest in data-driven methods for modeling, analysis, and control of such systems, which are otherwise not easily described by insightful physics-based models. While many purely data-driven black box modeling techniques have existed, the past decade has seen increased interest in operator-based methods. Associated with a dynamical system, the Koopman and Perron Frobenius operators propagate observables and densities in a function space, respectively. When viewed differently, the operators propagate certain statistical or probabilistic features of the dynamical system discussed in \cite{lasota_1994}, which can be sampled through measurements. Data-driven methods based on the operator methods therefore are expected to preserve a link to the physics of the system and provide insights while being a mathematical tool to create linear models. 

The Koopman operator propagates observables in time and therefore is particularly attractive to create approximate state observers. At the same, the Koopman operator can be used to create reduced order models, as in the early works on Dynamic Mode Decomposition (DMD) in \cite{mezic_jfm_2009, schmid_jfm_2010,h_tu_dynamic_2014}.  DMD and its cousins, such as Extended Dynamic Mode Decomposition (EDMD), rely on having a large data set of observables, with the precision of the approximate models improving in general with the amount of data available. These methods rely on computational steps such as the singular value decomposition and matrix inversion, which become infeasible as the amount of data increases. To overcome these challenges, some recent papers have suggested updating the Koopman operator using recursive least square computation in DMD calculations in \cite{zhang2019online} and in EDMD with forgetting factor \cite{csow_ecc_2021}, where the Koopman update was based on prediction error. However, there exists another particular problem for streaming data i.e., when data is collected at a high frequency over a prolonged period (or indefinitely) and when such data can intermittently be drawn from very different dynamical regimes of the system. Existing methods have to necessarily retain a historical data archive and as new data streams in, add to the existing archive, thereby presenting challenges to computational resources as shown in the early works for DMD without control in \cite{Hemati_2014} where the streaming dataset was used to update the DMD operator based on the residual of the Gram-Schmidt process. On the other hand, adaptive methods with a forgetting factor can ignore older data drawn intermittently from different dynamical regimes, regimes which may be visited by a trajectory again at a future time. Furthermore, updating the DMD operator with control in \cite{Hamdan_2023} using the subspace tracking method where the subspace of the observable space is being tracked with an assumption of a known system order. Intermittently different data or data from rare events is important to model and control for what could be a critical phenomenon, and this will require computationally efficient updates not only to the Koopman operator but also to determine the system order and learn the basis functions for an unknown complex system dynamics.

The essential idea underlying this paper is that observation data from a nonlinear dynamical system can be approximately represented as a linear high dimensional (of unknown dimension) time-invariant system using the Koopman operator and that this LTI system can be identified using the ideas from the recursive subspace system identification (R-SSID). In the proposed framework, a subspace representing the extended observability matrix of the high-dimensional LTI system is updated using the streaming data. Firstly, the extended observability matrix and the Koopman operator are approximated with an initial batch of data. As more data streams in, a check is performed on the Grassmannian distance between the subspace of the extended observability matrix and the new data segment. If this distance is more than a specified threshold, the Koopman operator is updated, and the segment of the streaming data is added to the archive; otherwise, the data is discarded. We demonstrate through examples that this online update process identifies streaming data from new dynamical regimes, and the updated Koopman operator can retain high predictive accuracy. We also identify the singular value-based minimum system order and learn the basis functions using Gaussian process regression.

The remainder of the paper is organized as follows. In sections \ref{II-A} and \ref{II-B}, we first introduce the fundamental concepts of the Koopman operator theory and an overview of the subspace system identification (SSID) algorithm. In section \ref{sec:recursive}, we introduce the R-SSID algorithm, and in section \ref{II-D}, the Gaussian process as a lifted observable is shown. In section \ref{III}, we introduce the proposed algorithm, which is the R-SSID with Grassmannian-guided data pruning. In section \ref{IV}, we introduce simulated examples to show the efficacy of the algorithm against EDMD in terms of prediction accuracy and end the paper with a conclusion in section \ref{V}.

\section{KOOPMAN OPERATOR ESTIMATION VIA SUBSPACE IDENTIFICATION}\label{II}

\subsection{Koopman operator}\label{II-A}
Consider a discrete-time dynamical system defined as,
\begin{equation*}
    x_{t+1} = F(x_t)
\end{equation*}
where $t$ is the time instant, $x_t$ is the state space that resides in the manifold $\M \subseteq \RR^{d} $ and $F:\M \mapsto \M$ represents the mapping between successive time steps. Given the function space $\G$ comprising of all functions that maps $\M \mapsto \RR$ referred to as observables $g \in \G $, the Koopman operator $\K$ acts on the observable as follows,
\begin{equation}
    \K g(x_t) = (g\circ F)(x_t) = g(x_{t+1}).
    \label{eq:koopman_def}
\end{equation}
 Eq.\eqref{eq:koopman_def} defines how the Koopman operator propagates the observable function $g$ along the state trajectories in the discrete-time dynamical system, see \cite{lasota_1994} or \cite{mezic_chaos_2012} for a review.

\begin{figure}[ht]
    \centering
    \includegraphics[width=0.75\columnwidth]{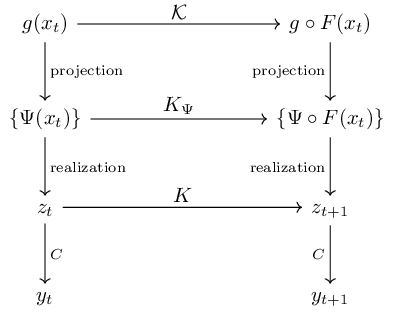}
    \caption{Commutative diagram illustrating the propagation of the observable $g(x_t)$, projected observable $\Psi(x_t)$ and its realization $z_t$ using the operators $\K, K_{\Psi}$ and $K$, respectively. }
    \label{fig:Koopman_diagram}
\end{figure}

Approximating the Koopman operator as a finite dimensional operator involves projecting the operator's action onto a finite-dimensional subspace, often called the 'lifted space.' This has been explored through various approaches such as basis function expansion, which can be defined a priori see \cite{williams2015edmd} and \cite{korda_mezic_2018} or learned through data-driven methods such as neural networks, see \cite{kevrekidis_chaos_2017}, \cite{lusch2018deep} and \cite{takeishi2018learning}. In this paper, we consider the sequence of measurements $y_t \in \RR^p$ as an evaluation of an observable function of the state, which we represent as a linear combination of finite basis functions denoted by $\Psi(x_t)$. This basis function evolves through an approximation of the Koopman operator $K_{\Psi}$ as 
\begin{equation}
    \K g(x_t) \approx \K [c^{\intercal} \Psi](x_t) \approx (K_{\Psi}c)^{\intercal} \Psi(x_t).
    \label{eq:koopman_proj}
\end{equation}
To derive an approximated set of dictionary function $\Psi$, we initially obtain a lifted state $z_t \in \RR^r$ and the corresponding operator $K$, see fig.\ref{fig:Koopman_diagram}. This process employs subspace identification (SSID) methods using temporal data sequence, bypassing the need to approximate $K_{\Psi}$ or specify the basis functions $\Psi$. 
In this approach, we model the regression or mapping between the states $x_t$, and the lifted states $z_t$ as a Gaussian process. 
\begin{equation}
    z_t \sim \Psi(x_t) = \mathcal{GP}(\mu(x_t),k(x_t,x_t))\,.
\end{equation}
The lifted states $z_t$ can then be viewed as a scalar-valued realization of a random variable in a function space evaluated for a particular $x_t$. This provides a probabilistic framework for the Koopman operator to propagate the observables as a function of time without first specifying or learning a dictionary of basis functions. 

For a discrete controlled system of the form 
\begin{equation}\label{eq:nonlinear_control}
    x_{t+1} = F_u (x_t,u_t)
\end{equation}

where $u_t \in \U \subset \RR^m$ is the control input and $F_u: \M \times \U \mapsto \M$ is the mapping between successive time steps for the controlled system. We take a similar approach in identifying the system matrices $(K,B,C,D)$ of lifted dynamics using subspace identification methods and leverage Gaussian process regression for mapping states to the lifted space. The formulation is then extended to a linear control system on the lifted state $z_t$ with a control input $u_t$ at time $t$ assumed to be
\begin{subequations}\label{eq:linear}
\begin{align}
z_{t+1}  & = Kz_t + Bu_t,  \\
y_t & = C z_t + D u_t.
\end{align}
\end{subequations}
It should be emphasized that selecting this model structure involves balancing simplicity for practical controller designs with the model's accuracy in real nonlinear systems, where LTI controllers may falter with substantial modeling errors against the choice of Bi-linear system \cite{DGoswami_Bilinear2021}.

\subsection{An overview on subspace identification (SSID)}\label{II-B}
In this subsection, the subspace identification method for identifying the system in eq.\eqref{eq:linear} using multiple data records is briefly reviewed. A linear time-invariant, noise-free discrete system described in eq.\eqref{eq:linear} is to be identified using the input-output data $\D_i = \{u_{t}^i, y_{t}^i\}_{t=0}^{n}$ for $i = 1,\dots, N.$, where $i$ is the number of data records and $t$ represents the time step for each data record. We assume that the system $(K,C)$ is observable but make no direct assumption about the order $r$ of the system. The Hankel matrices for each data record consisting of $n+1$ measured I/O data can be described with depth $l$ and length $s = n-l+1$ with a condition to have sufficient columns such that $s > lm+r$, see \cite{holcomb2017subspace}, 
\setlength\arraycolsep{1pt}
\begin{equation}
    \begin{split}\label{eq:block_Hankel}
        Y_i =
        \begin{pmatrix}
            y_{0}^i & y_{1}^i & \cdots & y_{{s-1}}^i\\
            y_{1}^i & y_{2}^i & \cdots & y_{{s}}^i\\
            \vdots  & \vdots    & \ddots & \vdots\\
            y_{{l-1}}^i & y_{{l}}^i & \cdots & y_{{n}}^i
        \end{pmatrix},~~
         U_i =
        \begin{pmatrix}
            u_{0}^i & u_{1}^i & \cdots & u_{{s-1}}^i\\
            u_{1}^i & u_{2}^i & \cdots & u_{{s}}^i\\
            \vdots  & \vdots    & \ddots & \vdots\\
            u_{{l-1}}^i & u_{{l}}^i & \cdots & u_{{n}}^i
        \end{pmatrix}.
    \end{split}
\end{equation}
From eq.\eqref{eq:linear} it follows,
\begin{equation}\label{eq:concat_single}
    Y_i = \Gamma_l Z_{0,i}^{\mathrm{lift}} + \H_l U_i
\end{equation}
where,
\begin{equation}
    \begin{split}
        \Gamma_l = \begin{bmatrix}
            C \\
            CK \\
            CK^2\\
            \vdots \\
            CK^{l-1}
        \end{bmatrix},
        \H_l = \begin{bmatrix}
            D & 0 & 0 & \cdots & 0 \\
            CB & D & 0 & \cdots & 0 \\
            CKB & CB & D & \cdots & 0 \\
            \vdots & \vdots & \vdots & \ddots & \vdots \\
            CK^{l-2}B & \cdot & \cdot & \cdots & D
        \end{bmatrix}.
    \end{split}
\end{equation}
where $\Gamma_l \in \RR^{lp\times r}$ is the extended observability matrix, $\H_l \in \RR^{lp \times lm}$ is the lower triangular block-Toeplitz matrix and  $Z_{0,i}^{\mathrm{lift}} \in \RR^{r\times s}$ is the realization of the lifted state that corresponds to the initial condition of each column trajectory for the dataset $i$, in the original state space. This system representation in eq.\eqref{eq:concat_single} can also be extended for multiple data records by simply appending the new I/O data Hankel matrices to the previous one to form mosaic-Hankel matrix as,
\begin{equation}\label{eq:mosaic-Hankel}
\begin{split}
     \mathbf{Y}_N =& \begin{bmatrix}
        Y_1, & Y_2, & \cdots, & Y_N
    \end{bmatrix} \in \RR^{l p \times Ns}\\
    \mathbf{U}_N =& \begin{bmatrix}
        U_1, & U_2, & \cdots, & U_N
        \end{bmatrix} \in \RR^{l m \times Ns}.
\end{split} 
\end{equation}
Now, arranging eq.\eqref{eq:concat_single} from dataset $i=1~to~N$ it gives us 
\begin{equation}\label{eq:concat_multi}
    \mathbf{Y}_N = \Gamma_l Z_{0,1:N}^{\mathrm{lift}} + \H_l \mathbf{U}_N.
\end{equation}
where $Z_{0,1:N}^{\mathrm{lift}} \in \RR^{r\times Ns}$ is the concatenation of $Z_{0,i}^{\mathrm{lift}}$ for datasets $i=1 ~ to~N$. Now, to find the column space of the extended observability matrix, we project out the influence of the input matrix from eq.\eqref{eq:concat_multi} by post-multiplication of the projection matrix defined as 
$\Pi_{\mathbf{U}_N}^{\perp} = I - \mathbf{U}_N^{\intercal}(\mathbf{U}_N \mathbf{U}_N^{\intercal})^{-1}\mathbf{U}_N$ which gives us,
\begin{equation}\label{eq:concat_without_u}
    \mathbf{Y}_N \Pi_{\mathbf{U}_N}^{\perp} = \Gamma_l Z_{0,1:N}^{\mathrm{lift}} \Pi_{\mathbf{U}_N}^{\perp}.
\end{equation}
The left side of the eq.\eqref{eq:concat_without_u} is available to us as it comprises of I/O data. As its column space coincides with that of $\Gamma_l$, we can take singular value decomposition of $\mathbf{Y}_N \Pi_{\mathbf{U}_N}^{\perp}$ 
\begin{equation}
    \mathbf{Y}_N \Pi_{\mathbf{U}_N}^{\perp}  = 
    \begin{bmatrix}
    Q_r ,~ Q_{l-r} 
    \end{bmatrix} 
    \begin{bmatrix} 
    \Sigma_r & 0 \\ 0 & \Sigma_{l-r} \end{bmatrix} 
    \begin{bmatrix}
            V_r^{\intercal} \\ V_{l-r}^{\intercal} 
            \end{bmatrix}.
\end{equation}
As $\Sigma_r$ contains the $r$ largest singular values, we approximate $\Gamma_l \approx \Gamma_r$ and determine the order of the lifted dynamics to be '$r$'. Therefore, the extended observability matrix is given as 
\begin{equation}\label{eq:reduced_subspace}
    \Gamma_r = Q_r \Sigma_r^{\frac{1}{2}}.
\end{equation} 
After computing the extended observability matrix the system state space matrices ($K,B,C,D$) and also the lifted state realization $Z_{0,1:N}^{\mathrm{lift}}$ can also be computed using approximated subspace $\hat{\Gamma_r}$ and Hankel matrix $\mathbf{Y}_N$. For detailed procedure to compute system matrices please refer \cite{Verhaegen1992SubspaceMI} and for multiple data records, refer \cite{holcomb2017subspace,LBT_MECC23} and further investigation into different subspace algorithms, see \cite{overschee_1994}, \cite{QIN20061502}, 

\subsection{Recursive algorithm for subspace identification (R-SSID)} \label{sec:recursive} 

\begin{algorithm}[htpb]
    \caption{ \cite{Oku1999AR4} Suppose $\Xi_{N}:= \mathbf{Y}_N \Pi_{\mathbf{U}_N}^{\perp} \mathbf{Y}_N,~ P_N:= (\mathbf{U}_N \mathbf{U}_{N}^{\intercal})^{-1}$ and $\mathbf{Y}_N \mathbf{U}_{N}^{\intercal}$ have already been obtained at previous step. When $(N+1)^{th}$ dataset  $\D_{N+1} =  \{u_t^{N+1} , y_t^{N+1}\}_{t=0}^n $ is obtained, the $(N+1)^{th}$ symmetric data matrix $\Xi_{N+1}$ can be updated by the following procedure:}
    \begin{algorithmic}[1]
        \State \textbf{Sample:} $\D_{N+1} =  \{u_t^{N+1} , y_t^{N+1}\}_{t=0}^n $ \Comment{New Dataset}
        \State \textbf{Construct:} $Y_{N+1}, U_{N+1}$  \Comment{As in eq.\eqref{eq:block_Hankel} s.t $s > lm+r$} 
        \For{$k = 1,2,\dots, s$}
            \State $\mathbf{u}_k \gets U_{N+1}(:,k)$ \Comment{Matlab notation $(:,k)$}
            \State $\mathbf{y}_k \gets Y_{N+1}(:,k)$
            \State \textbf{Update:}
            \begin{subequations}\label{eq:R_SSID}
                \begin{align}
                & \alpha_{N+1} \gets \big(1+\mathbf{u}_k^{\intercal} P_N \mathbf{u}_k \big)^{-1} \label{eq:R_SSID1} \\
                & e_{N+1} \gets  \mathbf{y}_k - \mathbf{Y}_N \mathbf{U}_{N}^{\intercal} P_N \mathbf{u}_k \label{eq:R_SSID2} \\
                & \Xi_{N+1} \gets  \Xi_N + \alpha_{N+1}~ e_{N+1} e_{N+1}^{\intercal} \label{eq:R_SSID3} \\
                & P_{N+1} \gets  P_N - \alpha_{N+1} P_N \mathbf{u}_k           \mathbf{u}_k^{\intercal} P_N \label{eq:R_SSID4} \\
                & \mathbf{Y}_{N+1} \mathbf{U}_{N+1}^{\intercal} \gets  \mathbf{Y}_N \mathbf{U}_N^{\intercal} + \mathbf{y}_k \mathbf{u}_k^{\intercal} \label{eq:R_SSID5}
            \end{align}        
            \end{subequations}
                   
        \EndFor
        \State $[Q,\Sigma^2,V^{\intercal}] = svd(\Xi_{N+1})$
        \State $\Gamma_{N+1} = Q_r \Sigma_r^{\frac{1}{2}}$ \Comment{'r' dominant singular values}
        \State \textbf{Obtain:} $(K,B,C,D)$ from $\Gamma_{N+1}$ 
        \State $N \rightarrow N +1$ \Comment{Move to the next recursive computation}
\end{algorithmic}\label{alg_1}
\end{algorithm}

 

The update to $\mathbf{Y}_N \Pi_{\mathbf{U}_N}^{\perp}$ indirectly updates the subspace of the extended observability matrix $\mathbf{\Gamma}_{N}$, a critical element in constructing the quadruple system matrices $(K,B,C,D)$. Suppose a new dataset $\D_{N+1} =  \{u_t^{N+1} , y_t^{N+1}\}_{t=0}^n $ is obtained the new mosaic-Hankel matrices in eq.\eqref{eq:mosaic-Hankel} can be updated as,
\begin{equation}\label{eq:Updated_mosaicHankel}
   \mathbf{U}_{N+1} = \begin{bmatrix}
    \mathbf{U}_N ,& U_{N+1}
    \end{bmatrix}, \quad
    \mathbf{Y}_{N+1} = \begin{bmatrix}
    \mathbf{Y}_N ,& Y_{N+1}
    \end{bmatrix}.
\end{equation}
 From above eq.\eqref{eq:Updated_mosaicHankel}, we see that the new data matrix $\mathbf{Y}_{N+1} \Pi_{\mathbf{U}_{N+1}}^{\perp}$ is not suitable for a recursive update as its column size increases. Therefore, we define a real square symmetric matrix $\Xi_N$,
 \begin{equation}
 \begin{split}
     \Xi_N &= \mathbf{Y}_N \Pi_{\mathbf{U}_N}^{\perp} \Pi_{\mathbf{U}_N}^{\perp} \mathbf{Y}_N = \mathbf{Y}_N\Pi_{\mathbf{U}_N}^{\perp} \mathbf{Y}_N, \\
        & = \begin{bmatrix} Q_r ,~ Q_{l-r} \end{bmatrix} 
        \begin{bmatrix} \Sigma_r^2 & 0 \\ 0 & \Sigma_{l-r}^2 \end{bmatrix} \begin{bmatrix}  Q_r^{\intercal} \\ Q_{l-r}^{\intercal} \end{bmatrix}.
 \end{split}   
 \end{equation}
This representation of the data matrix helps in recursive updates and still preserves the subspace that helps determine the system matrices.

After updating the squared data matrices, the system matrices $(K, B, C, D)$ can be acquired with each update step by performing an eigenvalue decomposition of the updated symmetric data matrix $\Xi_{N+1}$ to compute the subspace $\mathbf{\Gamma}_{N+1}$. Here we adopt the notation that $\Gamma_i$ represents the dataset $i$ and the bold $\mathbf{\Gamma}_i$ represents the subspace spanned by extended observability matrix updated recursively using dataset $[1,2,..,i]$.  Additionally, dominant modes can be selected for the subspace to update the system order $r$. This enables us to update the Koopman operator and, when necessary, adjust the system order. For a comprehensive derivation of algorithm \ref{alg_1}, which is based on the matrix inversion lemma, please refer to \cite{Oku1999AR4}. This source illustrates that the algorithm is an extension of the recursive least squares method. 

\subsection{Gaussian process as lifted observables}\label{II-D}
A Gaussian process $\mathcal{GP}(\mu,k)$ is defined as a collection of random variables, one for each point in input space $\XX$, such that the finite subset of these random variables follows a multivariate Gaussian distribution. In other words, for any finite set of input points $\XX = \{x_1,x_2,x_3, \cdots x_{n_g}\}$, the Gaussian process defines a joint distribution over the corresponding function values $\Psi(\XX) \sim Z_{0,x}^{\mathrm{lift}}= \{Z_{0,x_1}^{\mathrm{lift}}, Z_{0,x_2}^{\mathrm{lift}}, Z_{0,x_3}^{\mathrm{lift}}, \cdots Z_{0,x_{n_g}}^{\mathrm{lift}} \}$ as follows,
\begin{equation}
    z_j|\D_{GP} \sim \mathcal{N} \Big(\mu_{z_j| \D_{GP}} ( \XX ) , k_{z_j | \D_{GP}} (\XX) \Big)
\end{equation}
where, $\D_{GP} = \{\XX,  Z_{0,x}^{\mathrm{lift}} \}$ is the data on which the Gaussian process is conditioned and $j = 1,2,\cdots,r$ is the number of lifted states. The input points $\XX$ contain the state measurements of $x_t$, which corresponds to the particular lifted state realization $Z_{0,x}^{\mathrm{lift}}$ obtained through subspace identification. The posterior mean and covariance for an input point $x$ is calculated as 
\begin{align}
    \mu_{z_j}(x) =& \mu(x) + K_{x\XX}(K_{\XX\XX} + \sigma_{n}^2 I )^{-1}Z^{\mathrm{lift}}_{0,x}(j,:)^{\intercal} \label{eq:mean}, \\
    k_{z_j}(x,x) =& K_{xx} - K_{x\XX}(K_{\XX\XX} + \sigma_n^2I )^{-1}K_{\XX x}\label{eq:Cov}.
\end{align}
Here, the kernel matrix $K_{XY}$ represents the cross-covariance between the two sets $X$ and $Y$ evaluated using the ARD (Automatic Relevance Determination) squared exponential kernel function and $\sigma_n^2$ represents the measurement noise variance. The kernel parameters can be optimized through log-likelihood maximization. Here this
is done using the MATLAB toolbox for the Gaussian process regression to fit each of the $r$ GPs independently.

\section{RECURSIVE UPDATE USING GRASSMANNIAN DISTANCE}\label{III}
The Grassmannian $\mathrm{Gr}(k,n)$ represents all possible $k$-dimensional linear subspaces of an $n$-dimensional vector space.  The Grassmannian distance is a mathematical measure of the dissimilarity between the two subspaces in a high-dimensional vector space. It is invariant under coordinate transformations and can be determined in relation to principal angles, which provide insights into the geometric relationship between subspaces. Then for subspaces $\mathbf{\Gamma_1, \Gamma_2} \in \mathrm{Gr}(r,lp)$, we form matrices $\Gamma_1,\Gamma_2 \in \RR^{lp \times r}$ whose columns are their respective orthonormal bases; then,
\begin{equation}\label{eq:Grass_dist}
   G = d_{\mathrm{Gr}(r,lp)}(\mathbf{\Gamma}_1, \mathbf{\Gamma}_2) = \Big[ \sum_{k=1}^r \theta_k^2 \Big]^{1/2}
\end{equation}
where, $\theta_k=cos^{-1}\big(\sigma_k(\Gamma_1^{\intercal}\Gamma_2)\big)$ is the principal angle between $\Gamma_1$ and $\Gamma_2$ and $\sigma_k$ gives the $k^{th}$ singular values of the matrix $\Gamma_1^{\intercal}\Gamma_2$. 

\begin{algorithm}
\caption{Recursive SSID with Grassmannian-Guided Data Selection}\label{alg:R_SSID_GrD}
\begin{algorithmic}[1]
\State \textbf{Initialize:} $ \Xi_i:= \mathbf{Y}_i \Pi_{\mathbf{U}_i}^{\perp} \mathbf{Y}_i~,~ P_i:= (\mathbf{U}_i \mathbf{U}_{i}^{\intercal})^{-1}$ and $\mathbf{Y}_i \mathbf{U}_i^{\intercal}$
\For{$i = 0,1 \dots, N$}
     \State \textbf{Sample:} $\D_{i+1} =  \{u_t^{i+1} , y_t^{i+1}\}_{t=0}^n $ \Comment{New Dataset}
     \State \textbf{Construct:} $Y_{i+1}, U_{i+1}$  \Comment{As in eq.\eqref{eq:block_Hankel} s.t.  $s > lm + r$}
     \State $\Pi_{U_{i+1}}^{\perp} \gets I - U_{i+1}^{\intercal} \Big(U_{i+1} U_{i+1}^{\intercal} \Big)^{-1} U_{i+1}$
     \State $[q,\sigma^2,v^{\intercal}] \gets svd\big(Y_{i+1} \Pi_{U_{i+1}}^{\perp} Y_{i+1}^{\intercal} \big)$
     \State $\hat{\Gamma}_{i+1} \gets q_r \sigma_r^{\frac{1}{2}}$
     \State $G \gets d_{Gr}(\mathbf{\Gamma}_i, \hat{\Gamma}_{i+1})$
     \If{$ G >  \epsilon $}
        \For{$k = 1,2,\dots, s$}
            \State $\mathbf{u}_k \gets U_{i+1}(:,k)$ \Comment{Matlab notation $(:,k)$}
            \State $\mathbf{y}_k \gets Y_{i+1}(:,k)$
            \State \textbf{Update:} \Comment{Using eq.\eqref{eq:R_SSID}}  \\ $\alpha_{i+1} , e_{i+1}, \Xi_{i+1},  P_{i+1} ~\& ~\mathbf{Y}_{i+1} \mathbf{U}_{i+1}^{\intercal}$ 
        \EndFor
        \State $[Q,\Sigma^2,V^{\intercal}] = svd(\Xi_{i+1})$
        \State $\Gamma_{i+1} = Q_r \Sigma_r^{\frac{1}{2}}$ \Comment{'r' dominant singular values}
    \EndIf  
\EndFor
\State \textbf{Obtain:} $(K,B,C,D)$ and $Z_{0,1:N}^{\mathrm{lift}}$ from $\Gamma_{N}$, $Y_{N}$
\State $z_j|\D_{GP} \sim \mathcal{N} \big(\mu_{z_j| \D_{GP}} ( \XX ) , k_{z_j | \D_{GP}} (\XX) \big)_{j=1}^r$
\end{algorithmic}
\end{algorithm}

The fundamental idea of the subspace methods is that the subspaces of projected data matrices can retrieve essential system dynamics; then, this concept of measurable distances to quantify similarity between subspaces can be used effectively to help filter out redundant datasets and optimize the number of datasets on which recursive identification is performed. When the subspace is updated recursively using the algorithm 2, the Grassmannian distance $(G)$ can be computed between the subspace $\mathbf{\Gamma}_i$ identified using previous datasets $\D_{1:i}$ and the subspace $\Gamma_{i+1}$ of the new dataset $\D_{i+1}$. An arbitrary threshold value $\epsilon$ such that when $G<\epsilon$ then the dataset $\D_{i+1}$ is a $\epsilon$-redundant dataset and if $G>\epsilon$ the dataset is used for the recursive update. The Grassmannian distance for equidimensional subspace can be extended to subspaces of different dimensions as shown in \cite{ye2016schubert}. Therefore, even when system order $r$ is changed in algorithm \ref{alg:R_SSID_GrD}, the Grassmannian distance can still be used to differentiate between the identified subspaces. 

\section{SIMULATIONS AND RESULTS}\label{IV}
\subsection{Numerical example with simple Koopman embedding}
In this section, we demonstrate the proposed algorithm \ref{alg:R_SSID_GrD} in identifying the Koopman operator. Let us consider a continuous-time nonlinear system without control input:
\begin{equation}\label{eq:ex1_non}
\begin{split}
        \dot{x}_1  &= \mu x_1 \\ \dot{x}_2 &= \lambda (x_2 -x_1^2) \\  y &= [x_1, x_2]^{\intercal}.
\end{split}
\end{equation}
where, parameters $\lambda = -0.8$ and $\mu = -0.3$. 
For each dataset $\D_i$ ($i=0,1 \dots N$), we simulate the nonlinear system in eq.\eqref{eq:ex1_non} with a time step of $\delta = 0.1s$ and the number of time steps is $n=15$. First, the algorithm \ref{alg:R_SSID_GrD} is initialized and then, we stream $N=100$ datasets into the algorithm successively, where it constructs a Hankel matrix of an incoming dataset, with depth $l = 10$ and the number of columns $s = 6$. 
\begin{figure}[ht]\label{fig:EX1_train}
   \begin{subfigure}[b]{0.24\textwidth}
         \centering
         \includegraphics[width=\textwidth]{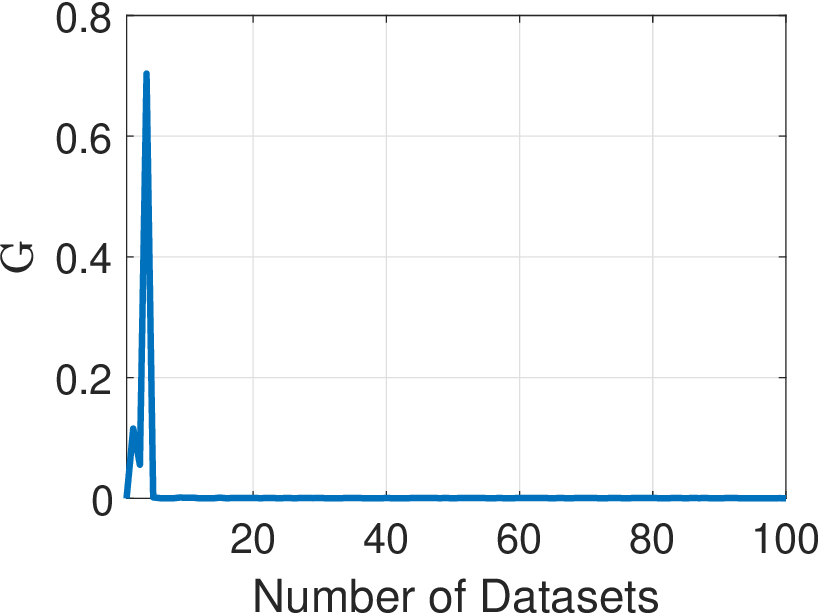}
         \caption{}
         \label{fig:EX1_Gr_r}
     \end{subfigure}
     \begin{subfigure}[b]{0.24\textwidth}
         \centering
         \includegraphics[width=\textwidth]{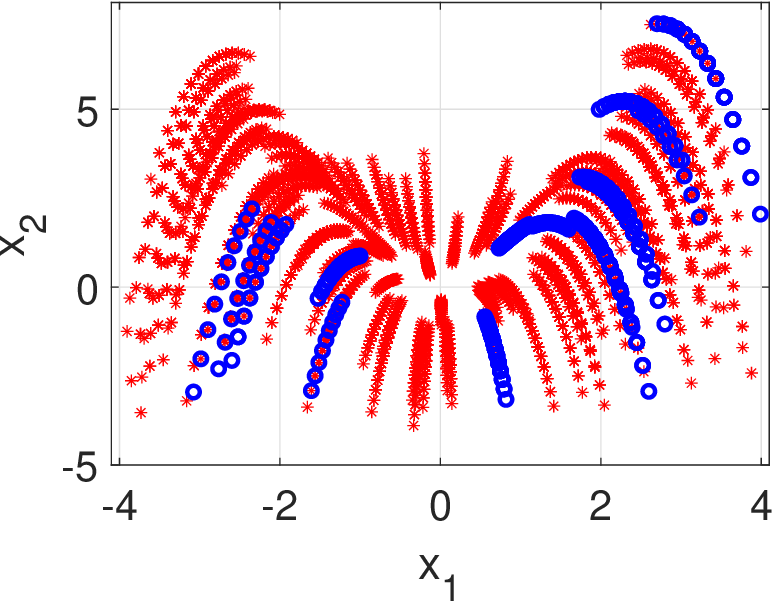}
         \caption{}
         \label{fig:EX1_TrainData}
     \end{subfigure}
     \caption{(a) Grassmannain distance $(G)$ between the subspace spanned by the extended observability matrix of the incoming dataset $\D_{i+1}$ and the recursively updated subspace $\mathbf{\Gamma_i}$. (b) The red points represent the $\epsilon$-redundant datasets, while the blue points represent the useful ones.}
\end{figure}
Here, in fig.\ref{fig:EX1_TrainData} the circular (blue) points are the data that is used in the recursive subspace identification, and the star (red) points are the $\epsilon$-redundant dataset based on the Grassmannian distance ($G$) and the $\epsilon=10^{-3}$. In this example, only 10 out of the 100 datasets are accepted for updating the subspace $\mathbf{\Gamma}_i$ as shown in fig.\ref{fig:EX1_Gr_r}.

\begin{figure}[ht]
   \begin{subfigure}[b]{0.24\textwidth}
         \centering
         \includegraphics[width=\textwidth]{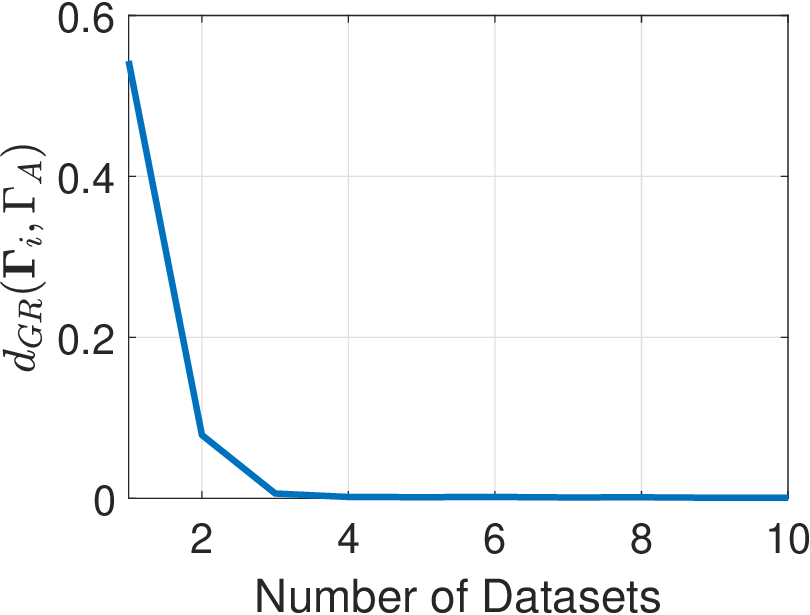}
         \caption{}
         \label{fig:EX1_Gr_D}
     \end{subfigure}
     \begin{subfigure}[b]{0.24\textwidth}
         \centering
         \includegraphics[width=\textwidth]{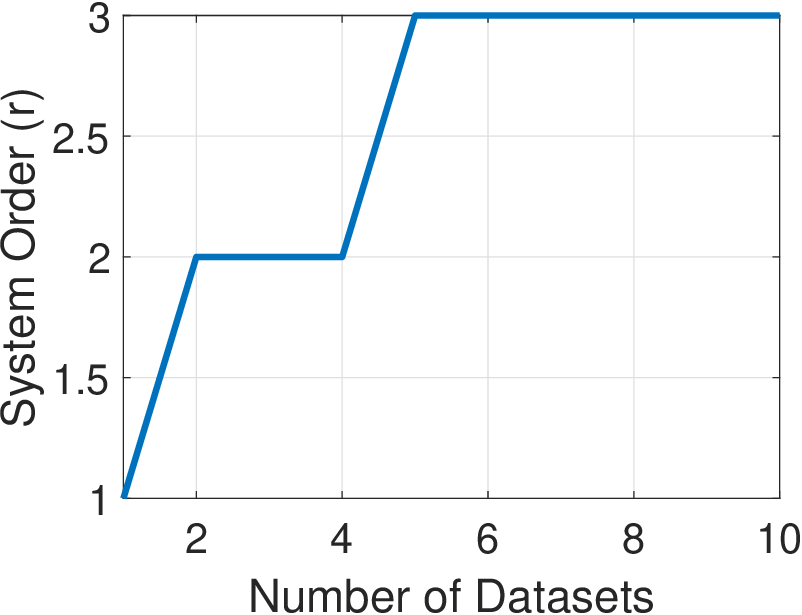}
         \caption{}
         \label{fig:EX1_oreder}
     \end{subfigure}
    \begin{subfigure}[b]{0.24\textwidth}
         \centering
         \includegraphics[width=\textwidth]{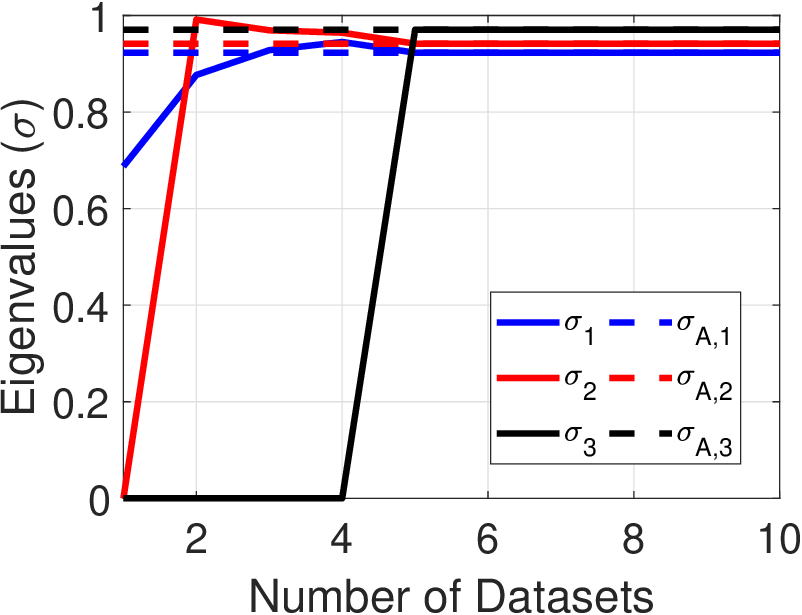}
         \caption{}
         \label{fig:EX1_eigenvalues}
     \end{subfigure}
     \begin{subfigure}[b]{0.24\textwidth}
         \centering
         \includegraphics[width=\textwidth]{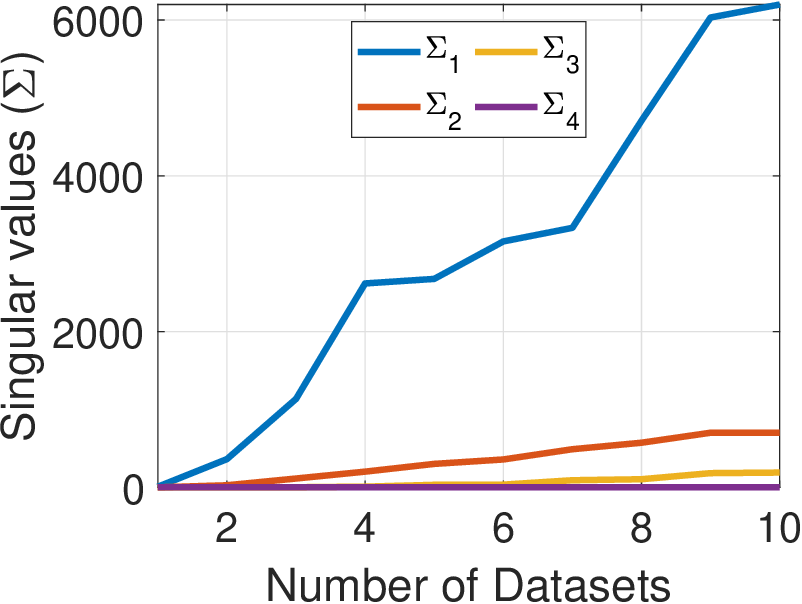}
         \caption{}
         \label{fig:EX1_singularvalues}
     \end{subfigure}
        \caption{(a) Grassmannian distance $d_{Gr}(\mathbf{\Gamma}_i, \hat{\Gamma}_{i+1})$ at each dataset between the subspace spanned by extended observability matrix of the system in eq.\eqref{eq:ex1_lin} and the identified system ($K,C$). (b) Identified System order ($r$). (c) Eigenvalues of $K_i$, updated with each streaming dataset. (d) Singularvalues of $\Xi_i$, updated with each streaming dataset.} 
        \label{fig:EX1}
\end{figure}

The state space of the nonlinear system in eq.\eqref{eq:ex1_non} can be expanded into a 3-dimensional linear system using nonlinear observations of state $x_1$ as $z_3 = x_1^2$, which is discussed in detail in \cite{brunton2021modern}. The linear discrete approximation of the system in eq.\eqref{eq:ex1_non} with the coordinates $[z_1, z_2, z_3]^{\intercal} = [x_1, x_2, x_1^2]^{\intercal}$ in the form of $z(t+1) = A z(t)$ is given as 
\begin{equation}\label{eq:ex1_lin}
\begin{split}
        z(t+1)  = &
    \begin{bmatrix}
        0.9704 & 0 & 0\\ 0 & 0.9231 & 0.0746 \\ 0 & 0 & 0.9418
    \end{bmatrix}
        z(t)  \\ 
    y(t) =& ~[z_1(t), z_2(t)]^{\intercal}
\end{split}
\end{equation}
The system in eq.\eqref{eq:ex1_lin} represents the true discrete Koopman model for the nonlinear system in eq.\eqref{eq:ex1_non}, and therefore we use this system to compare our identified linear system. Fig.\ref{fig:EX1_Gr_D} shows the Grassmannian distance between the subspace of recursively updated $\mathbf{\Gamma}_i$ and the subspace spanned by the extended observability matrix of the system in eq.\eqref{eq:ex1_lin}, represented as $\Gamma_A$. We observe that the Grassmannian distance decreases to zero, which shows the similarity between the extended observability matrix of the two systems. Fig.\ref{fig:EX1_eigenvalues} illustrates a continuous update of the eigenvalues $(\sigma)$ of the estimated Koopman operator $K$, which gradually approaches the eigenvalues $(\sigma_A)$ of the LTI system in eq.\eqref{eq:ex1_lin}. Hence, the identified system matrix $K$ is similar to the true Koopman model matrix $A$. 
Fig.\ref{fig:EX1_singularvalues} shows the increase in the magnitude of singular values of the square data matrix $\Xi_i$ as more data is incorporated, and the number of dominant singular values $(\Sigma)$ indicates the order of the system. From fig.\ref{fig:EX1_oreder}, the order of system in eq.\eqref{eq:ex1_non} can be seen to be 3, as fourth singular value is almost zero.

\subsection{Duffing oscillator with control input}
\begin{figure}[ht]
    \centering
    \includegraphics[width=0.75\columnwidth]{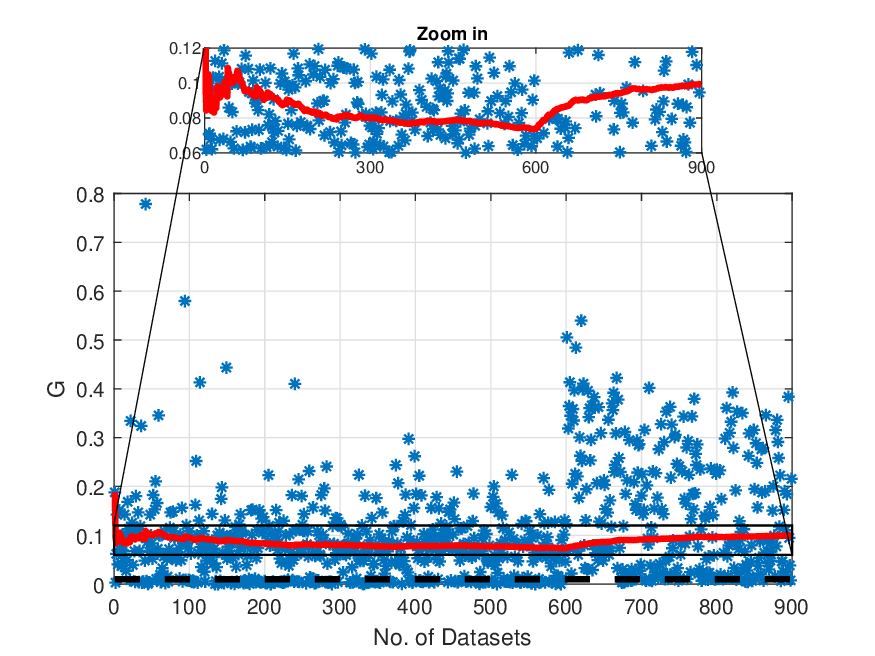}
    \caption{Grassmannain distance $(G)$ between the subspace $\Gamma_{i+1}$ spanned by the extended observability matrix of the incoming dataset and the recursively updated subspace $\mathbf{\Gamma_i}$. The dotted black line represents the threshold value of $\epsilon=0.01$. The red line represents the cumulative average of $G$ computed over the number of datasets.}
    \label{fig:Grassman_Duf}
\end{figure}

 This subsection examines the proposed algorithm against an example that does not lend itself to a straightforward Koopman embedding but exhibits a wide range of complex behaviors, such as the Duffing oscillator with a control input given as   
\begin{equation}\label{eq:Duffing}
\begin{split}
        \dot{x}_1  &= x_2 \\ \dot{x}_2 &= \alpha x_1 + \beta x_1^3 + u \\  y &= [x_1, x_2]^{\intercal}.
\end{split}
\end{equation}
where $\alpha = -\beta = 1$ and the control input $u \in \{-0.5,0.5\}$. To identify the Duffing oscillator described in eq.\eqref{eq:Duffing} using the data-driven algorithm outlined in algorithm \ref{alg:R_SSID_GrD}, we continuously simulate and stream a dataset $\D_i= \{u_t^{i+1} , y_t^{i+1}\}_{t=0}^n$, with time-step $\delta = 0.01$ and number of time steps $n=800$. In processing the dataset, we construct a Hankel matrix with a depth of $l=200$ and a column length of $s=601$, ensuring that $s>ml+r$. Here, the threshold distance $\epsilon$ is chosen to be 0.01. In total, we stream $N=900$ datasets, out of which $22.22\%$ are rejected. Fig.\ref{fig:Grassman_Duf} shows the Grassmannian distance at each consecutive dataset as they were streamed.  

\begin{figure}[ht]\centering
         \includegraphics[width=0.45\columnwidth]{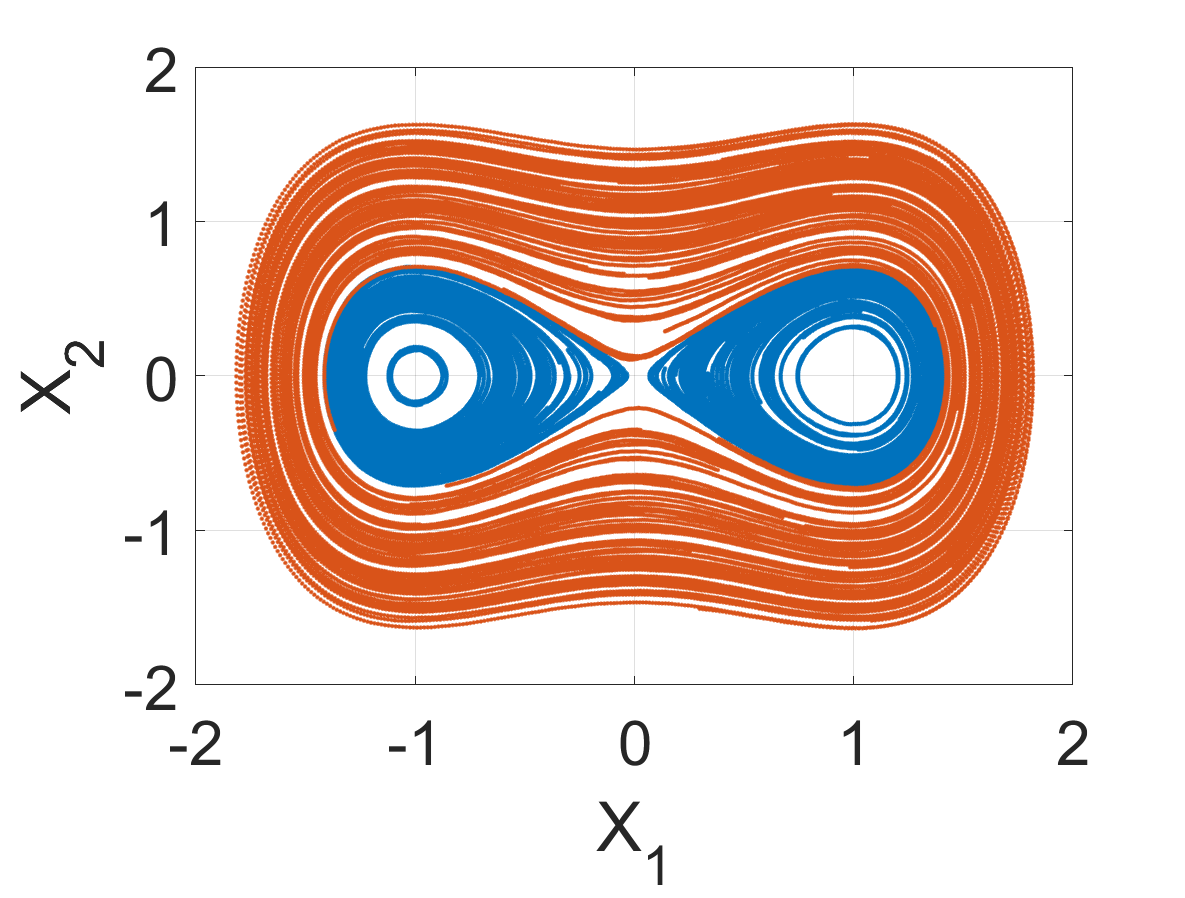}
         \caption{Trajectories sampled in the first 600 datasets are shown in blue (single-well oscillations), and the following datasets from 601 to 900 are shown in red (double-well oscillations).}
         \label{fig:data_duffing}
    \end{figure}

The zoomed-in portion of fig.\ref{fig:Grassman_Duf} highlights a notable increase in the cumulative mean of the Grassmannian distance at iteration $601$. This increase suggests that datasets containing previously unexplored subspaces are streamed. It's important to note that for the first 600 datasets, we selected initial conditions near two stable equilibrium points, giving us single-well oscillations as shown by blue trajectories fig.\ref{fig:data_duffing}. However, for the subsequent datasets, the initial points were chosen away from these equilibrium positions, which gave us double-well oscillations corresponding to red trajectories in fig.\ref{fig:data_duffing}. Therefore, in the context of the Duffing oscillator, qualitative difference in the single-well and double-well trajectories is expressed here by the sudden rise (at the dataset 601) in the cumulative average of Grassmannian distance between the subspaces spanned by the extended observability matrix of the new datasets (drawn from the inter-well oscillations shown in red in  fig.\ref{fig:data_duffing}) and previously identified system. 

    \begin{figure}[ht]\centering
         \includegraphics[width = 0.45\columnwidth]{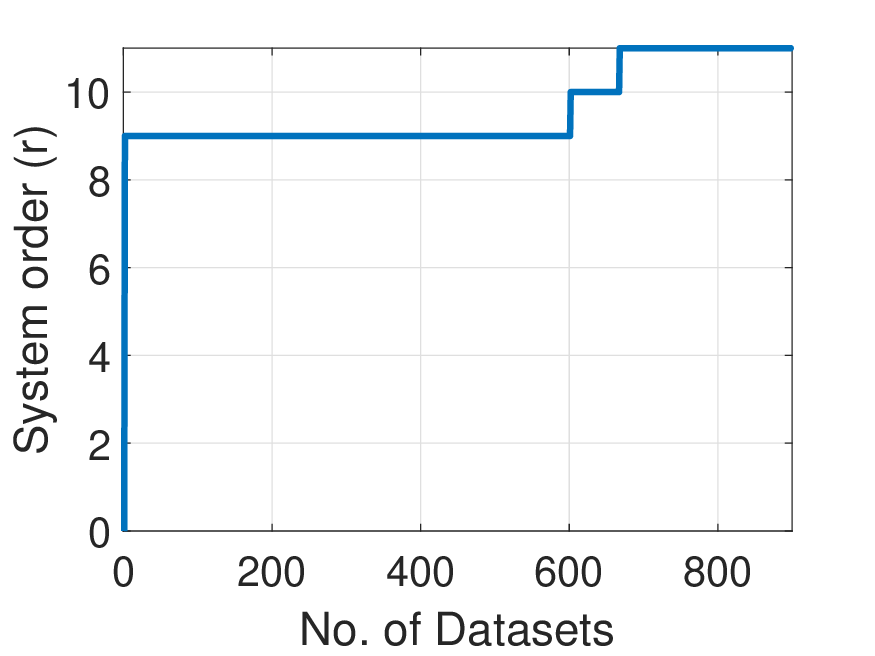}
         \caption{Identified System order (r) along the streamed dataset.}
         \label{fig:order_Duf}
    \end{figure}

In fig.\ref{fig:order_Duf}, it is evident that when algorithm \ref{alg:R_SSID_GrD} encounters information or a dataset whose subspace significantly deviates from the subspace spanned by the extended observability matrix of the previously updated system, it demonstrates the ability to adapt by increasing the order of the system to incorporate the new information. After processing a total of $N=900$ datasets, we utilize the updated subspace $\mathbf{\Gamma}_N$ in conjunction with $\mathbf{Y}_N$ to derive the system matrices $(K, B, C, D)$ and $Z_{0,1:N}^{\mathrm{lift}}$. We then employ Gaussian process regression in mapping the initial conditions of some trajectories in $\mathbf{Y}_N$ to $Z_{0,1:N}^{\mathrm{lift}}$. This step enables the lifting of the original states to the higher-dimensional lifted states for the identified linear lifted dynamics.

\begin{figure}[htpb]
   \begin{subfigure}[b]{0.24\textwidth}
         \centering
         \includegraphics[width=\textwidth]{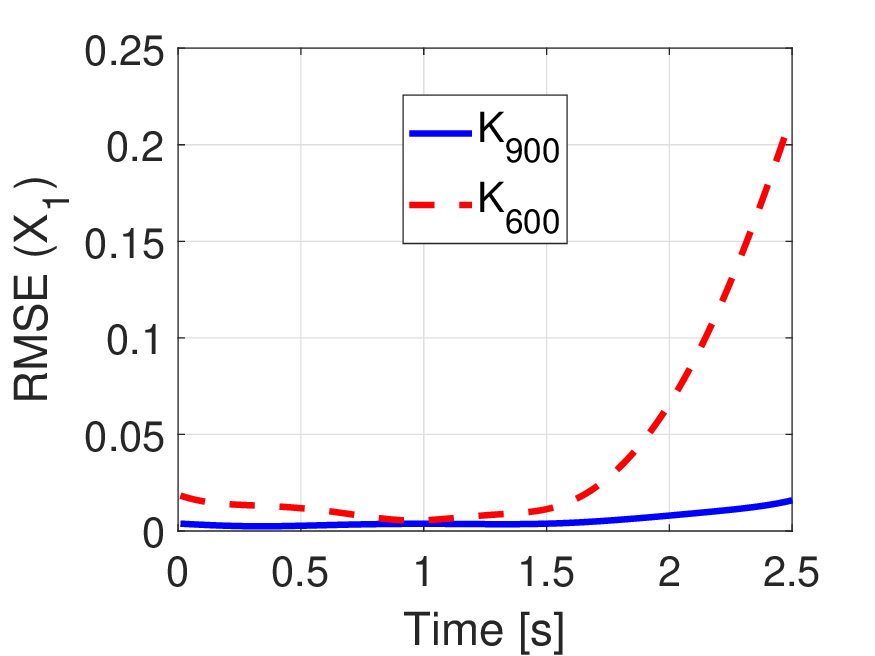}
         \caption{}
         \label{fig:RMSE_x2}
     \end{subfigure}
     \begin{subfigure}[b]{0.24\textwidth}
         \centering
         \includegraphics[width=\textwidth]{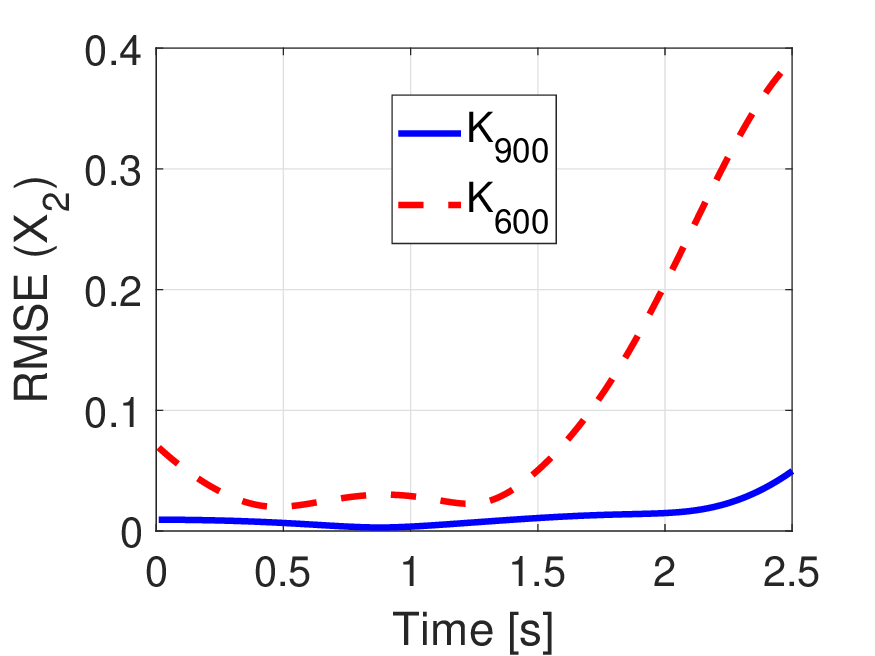}
         \caption{}
         \label{fig:RMSE_x1}
     \end{subfigure}
     \caption{Comparison of root mean squared prediction error for the system is $K_{600}$ (dotted red line) and  $K_{900}$ (solid blue line) for $x\in [-1.5,1.5]$ and randomly applying $u \in [-0.1,0.1]$ }\label{fig:RMSE_Duf}
\end{figure}

\begin{figure}[htpb]
   \begin{subfigure}[b]{0.24\textwidth}
         \centering
         \includegraphics[width=\textwidth]{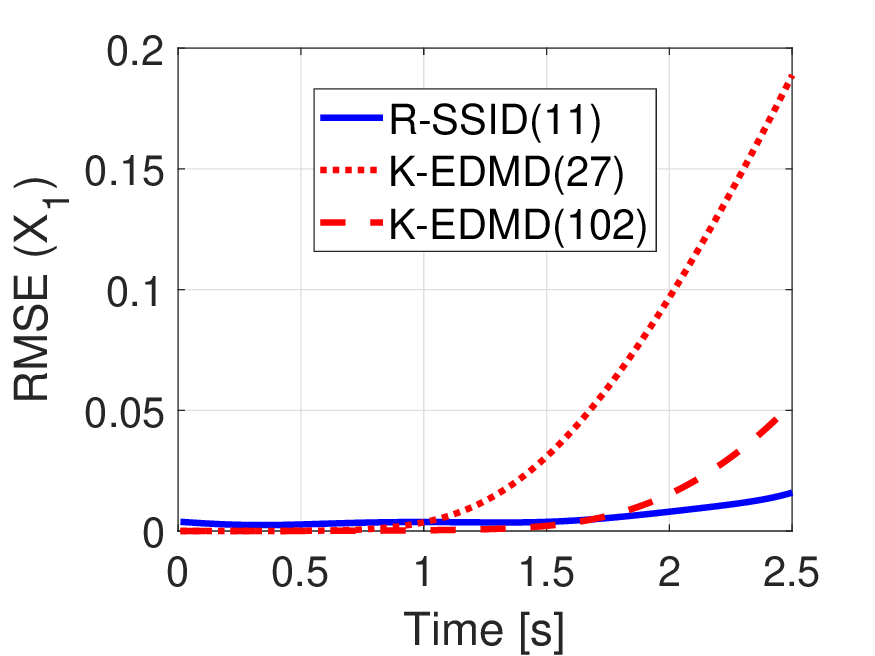}
         \caption{}
         \label{fig:RMSE_x2K}
     \end{subfigure}
     \begin{subfigure}[b]{0.24\textwidth}
         \centering
         \includegraphics[width=\textwidth]{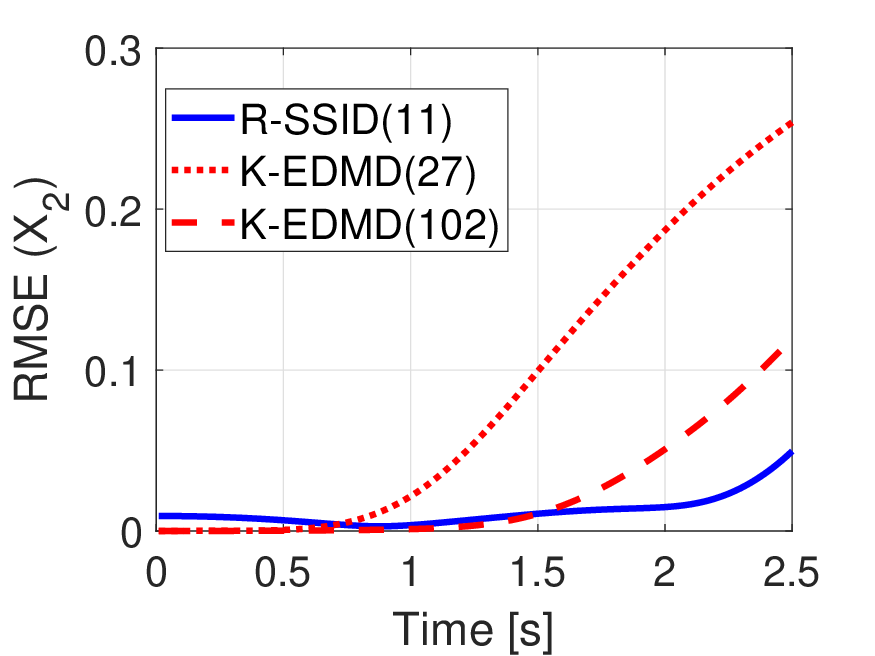}
         \caption{}
         \label{fig:RMSE_x1K}
     \end{subfigure}
     \caption{Comparison of root mean squared prediction error for R-SSID algorithm \ref{alg:R_SSID_GrD} (solid blue line), KEDMD with 27 lifting functions (dotted red line) and KEDMD with 102 basis functions (dashed red line)}\label{fig:RMSE_DufK}
\end{figure}

The system identified after streaming $600$ datasets is denoted as $K_{600}$ and has a system order of 9.  Meanwhile, the system identified after processing $900$ datasets is represented as $K_{900}$ and has a system order of 11. Fig.\ref{fig:RMSE_Duf} shows the root mean square prediction error for the system $K_{900}$ is less than the system $K_{600}$ computed over $100$ different initial conditions for 250-time steps. It shows the need for an algorithm that updates the Koopman operator as the data is streamed. In fig.\ref{fig:RMSE_DufK}, we compare the prediction error of our algorithm with the K-EDMD method with lifting functions as radial basis functions with centers chosen using the K-nearest neighbors algorithm and also the states themselves. We observe that our algorithm \ref{alg:R_SSID_GrD} (R-SSID) with $11$ lifting functions performs better than the K-EDMD with $27$ and $102$ lifting functions. The root mean squared prediction error is calculated similarly with $100$ different initial conditions and random control input $u$.      

The results conclude that algorithm 2 (R-SSID), firstly provides a considerable reduction in the basis functions and, secondly, exhibits lower prediction errors over a longer period compared to the EDMD method. We compare the recursive algorithm 2 with complete batch EDMD consisting of all the training data because it generates a similar Koopman operator as that of the online EDMD for the Linear Time-Invariant  (LTI) system. Notably, while prior studies, \cite{zhang2019online,Hamdan_2023}, highlight the necessity of updating the Koopman operator for Linear Time-Varying (LTV) systems, this paper focuses on a scenario involving data sampled from diverse dynamical regimes and a recursive algorithm is formulated to identify a reduced-order latent space which is represented through a high-dimensional LTI system. 

\section{CONCLUSION}\label{V}
This paper presents an algorithm that recursively updates the Koopman operator where the streaming data is available. The algorithm uses a recursive subspace identification method to compute the Koopman operator. This recursive update is made when the Grassmannian distance between the subspace of the extended observability matrix and the streaming segments of data is larger than a specified threshold. Hence different subspaces are identified with qualitatively different regimes in the system dynamics. In this method, the number of basis functions for the lifted space can change while the archive of data stored and used in computations is kept small. This is important in applications where storing streaming data over extended periods and manipulating very large data matrices can overwhelm computational resources. Our examples demonstrate that the proposed online update process accurately identifies new data patterns, maintains high predictive accuracy with the updated Koopman operator, and selects the minimum system order while learning basis functions through Gaussian process regression.





\bibliography{main}             
                                                   








\end{document}